# Positron elastic scattering by a semifilled-shell atom

M. Ya. Amusia,[1,2] V. K. Dolmatov,[3] and L. V. Chernysheva[2]

[1]*Racah Institute of Physics, Hebrew University, 91904 Jerusalem, Israel*
[2]*A. F. Ioffe Physical-Technical Institute, 194021 St. Petersburg, Russia*
[3]*University of North Alabama, Florence, Alabama 35632, USA*

We theoretically study the positron elastic scattering by an atom with a multielectron semifilled subshell in its structure. The positron scattering by the Mn($...3d^5 4s^2, ^6S$) atom with a $3d^5$ semifilled subshell ($e^+ +$Mn scattering) is chosen as a case study. We account for both the electron correlation and the formation of a $e^+ + e^-$ virtual positronium (Ps) in the intermediate states of the $e^+ +$ Mn system. Electron correlation is taken into account in the framework of the self-energy part of the scattering positron Green function generalized for the application to semifilled-shell atoms. The influence of the virtual Ps formation on $e^+ +$Mn scattering is taken into account by the reduction of the energy of the virtual *positron plus atomic-excited-configuration* states by the Ps-binding energy, to a reasonable approximation. We unravel the importance and specificity of the influence of both the virtual Ps and electron correlation on $e^+ +$ Mn elastic scattering. We demonstrate spectacular differences between the electron and positron scattering processes.

## I. INTRODUCTION

Positron-atom ($e^+ + A$) scattering is an interesting and important process. Its study has a long history (see, e.g., [1–11] and references therein).

Compared to electron-atom ($e^- + A$) scattering, the $e^+ + A$ scattering process is, at first glance, much simpler due to the lack of exchange between e$^+$ and the atomic electrons. Furthermore, qualitatively, the incident positron is subject to a rapidly increasing repulsion from a positively charged atomic nucleus as the atom is approached. This is in contrast to a stronger attraction of the incident electron to the nucleus of the target atom. As a result, one could expect that the cross section for scattering of an electron by an atom would be much larger than that for scattering of a positron at a given energy.

However, in contrast to a would-be simplicity of the $e^+ + A$ scattering, this process appears to be even more complicated than the electron scattering. This is due to the possibility of the formation of a $e^+ + e^-$ virtual positronium (Ps) between the incoming positron and a virtually excited electron of one of the atomic shells [1–4]. The latter is a multielectron process. This significantly complicates the description of the $e^+ + A$ scattering, because this process is an essentially multielectron process even without taking into account the formation of a virtual Ps. It turns out that this formation strongly affects the scattering of positrons by atoms to such an extent that, in some cases, the $e^+ + A$ elastic scattering cross section can even prevail over the $e^- + A$ scattering cross section.

Although many methods have been developed and attempts have been made to study $e^+ + A$ scattering with varying success, the main focus has been on positron scattering by inert and alkali atoms.

As far as we know, the scattering of positrons by atoms belonging to another group of atoms - the atoms with the highest spin multiplicity - has remained unexplored. These atoms have a multielectron semifilled subshell in their ground states, such as, for example, a $3d^5$ semifilled subshell in the Mn($...3d^5 4s^2, ^6S$) atom, which is a $3d$-transition-metal atom. Although studies were carried out on the interaction of positrons with a number of transition metal atoms [6], they were aimed at detection of bound states of the $e^+ + A$ system, rather than on elastic scattering of positrons by these atoms, at least not by transition-metal atoms with a semifilled multielectron shell. This has happened not because of a lack of interest in such a study, but because of the extreme difficulties of dealing with these atoms, both theoretically and experimentally. For example, the agreement between the only two currently existing experiments on elastic electron-Mn scattering [12, 13] leaves much to be desired. This calls for new experiments to resolve said disagreement. However, as far as we know, no such experiments have been carried out to date.

In this paper, in order to fill the gap in knowledge about the scattering of positrons by a multielectron atom with a semifilled shell, we choose Mn as a case-study-atom. The choice is not accidental. It is dictated by our rich experience in studying multielectron processes involving the Mn atom, such as various aspects of its photoionization (e.g., [14–16] and references therein) and elastic electron scattering (e.g., [17–19] and references therein). Moreover, the a priori results of our work [18] will give us a basis for a direct elucidation of the similarities and differences between the scattering of positrons and electrons by Mn. In addition, the choice of Mn for our case study may be of interest to related sciences as well. For example, to astrophysics, since Mn is found in abundance in various astrophysical objects. Therefore, knowledge of the collision of a positron with Mn can be useful for understanding astrophysical phenomena.

Positron scattering off a $3d$-transition-metal atom like Mn involves a number of important elements of novelty. Indeed, the $3d$ subshell of Mn has a kind of dual nature. On the one hand, it is an inner subshell, because it is collapsed into an inner region of the atom (its averaged radius is about 1.12 a.u.). On the other hand, this is a valence subshell in energy, because its ionization potential (approximately 14 eV) is close to the 8-eV ionization potential of the outermost $4s^2$ subshell. Earlier [20], the duplicity of the "inner-valence" $3d^5$ subshell was proven to be the key to the understanding of some differences between the $3p$-photoabsorption spectra of Mn, $Mn^+$, and $Mn^{2+}$. However, how such duplicity of the $3d$ orbital can affect positron scattering through a virtual Ps formation with a virtually excited $3d$-electron is not known. This presents an interesting novel topic for study. Furthermore, there are two distinctly different routes for the excitation of the $4s^2$ subshell of Mn(...$3d^5 4s^2$, $^6S$). One of them makes the ionic core of the excited Mn be a $4s^{-1}(^7S)$ state, whereas the other one results in a $4s^{-1}(^5S)$ state. The differences between the scattering of positrons during the formation of virtual Ps in different channels are unknown. Thus, clarifying these differences is another interesting topic for study. In addition, it is not clear a priori how the formation of a virtual Ps during the $e^+ +$ Mn scattering is affected by the interference between the two virtually excited channels of the $4s$ electrons, and how such "interference-formed" Ps does affect the scattering process itself. Neither is clear how both the formation of a virtual Ps and the scattering process are influenced by the interference between the virtually excited channels of the $4s^2$ subshell and the "inner-outer" $3d^5$ subshell of Mn.

Thus, certainly, the study of elastic positron scattering by high-spin semifilled-shell atoms is a topic of novelty and significance at least from a viewpoint of basic science. It is precisely the ultimate goal of the present paper to get insight into this topic.

In the present paper, we unravel how the perturbation of various atomic subshells of Mn by a scattering positron affects the scattering process both quantitatively and qualitatively, how this process is affected by the formation of virtual Ps in individual and coupled virtual states of the $e^+ +$ Mn system, and we reveal the similarities and differences between elastic $e^+ +$ Mn and $e^- +$ Mn scattering.

We calculate the correlated response of the $e^+ +$ Mn system to the scattering process in the framework of the self-energy part of the Green function of the incident positron. This is done similarly to the formalism of the Green function generalized to be applied to an atom with a semifilled shell in our earlier studies of the electron scattering [17, 18, 21].

To account for the virtual Ps formation upon $e^+ + A$ scattering is a difficult task overall. It requires to take into consideration the bound, exited and continuous states in the presence of the $e^+ + e^-$ pair in the field of the atom. This leads to the need to deal with at least the three-body problem which is not a simple task at all. Different types of approximations based on first principles have been suggested to solve the problem. However, not in all of them the virtual Ps formation is taken into account in a physically transparent manner. This is particularly typical for a close-coupling approximation (see, e.g., [7]). In the latter case, the Ps formation is to some extent taken into account by increasing the number of configurations taken into account in the calculations until reasonable agreement with experiment is obtained. However, the physical role of the phenomenon itself - the formation of Ps upon $e^+ + A$ scattering - is not visible in this approximation. In contrast, in our work, we aim at the use of a physically transparent approach to solving this problem. The most physically transparent theories, in our opinion, are the theories based on a combination of the Feynman diagram technique and the Dyson equation for the Green function of a scattering particle, such as methods developed, e.g., in [1, 3, 4, 8]. Among these methods, works [3, 8], particularly work [8], suggest significantly more complicated and complete methods than the method suggested in [1, 4]. Specifically, e.g., the approximation developed in [3] assumes that the relative motion of the $e^+ + e^-$ pair is determined by the real wavefunction of the ground-state Ps, whereas the motion of the center-of-mass of the virtual pair is considered unaffected by the atomic field; this is a drastic approximation to the real very complicated problem. As for work [8], it provides a well-developed, complete theoretical formalism, in which the electron-positron integral equation for Feynman ladder diagrams is converted into linear algebra equations by discretizing the electron and positron continuum states using a square-integrable basis of B-spline functions. However, one should recognize that the exact solution of this problem is not possible without subsequent approximations. In contrast, works [1, 4] suggest a simpler approximation to the scattering problem, in which the formation of Ps upon $e^+ + A$ scattering is accounted by reducing the virtual excitation energies of the atom by the magnitude of the actual Ps binding energy, $I_{\rm Ps} \approx 6.8$ eV. Despite such simplification is crude, it was proven to be a viable approximation. Indeed, in Fig. 1, we depict calculated data for $e^+ +$ Xe elastic scattering obtained in [4] and [8], and corresponding experimental data [22, 23].

One can see from Fig. 1 that a simple way to account for the formation of Ps upon $e^+ + A$ scattering, suggested in [1, 4], is, indeed, a viable approximation, albeit being cruder and less complete than that developed in [8]. Given that, in the present study, we aim to unravel the spectrum of effects that can occur in the characteristics of the $e^+ + A$ elastic scattering, exemplified by $e^+ +$ Mn scattering, as well as to reveal the main differences between the positron and electron scattering off Mn, rather than to carry out the most detailed calculation of the process, we use the approximation proposed and described in [1, 4].

Atomic units (a.u.) are used throughout the paper unless specified otherwise.

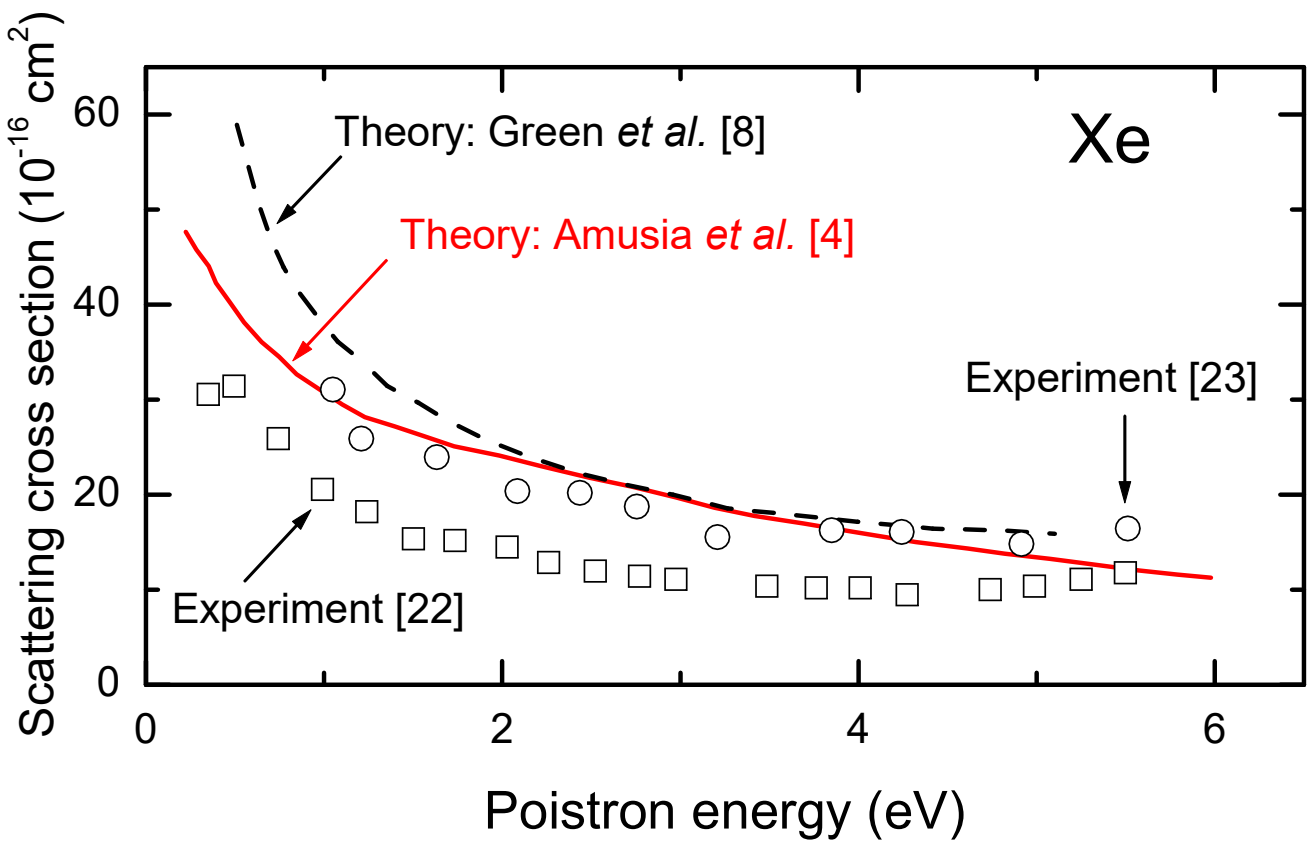

FIG. 1. Elastic $e^+ + \mathrm{Xe}$ scattering cross section. Solid, calculated data from Amusia *et al* [4]. Dash, calculate data from Green *et al* [8]. Open squares and circles, experimental data from Dababneh *et al* [22] and Sinapius *et al* [23], respectively.

## II. MAIN POINTS OF THEORY

The present study is based on the principles that were detailed in our previous calculations of the structure, photoionization of, and elastic electron scattering by, semifilled shell atoms [14, 15, 17, 18, 21, 24] (and references therein). Therefore, in the present paper, we only highlight the main provisions of the theory.

As the zeroth-order approximation, we choose the spin-polarized Hartree-Fock (SPHF) approximation [14, 24, 25]. In SPHF, an atom with a semifilled subshell in the ground state is considered as consisting of two different kinds of electrons - the "spin-up" (↑) electrons (a $z$-component of their spin is $s_z = +\frac{1}{2}$) and the "spin-down" (↓) electrons ($s_z = -\frac{1}{2}$). There is no exchange interaction between the spin-up and spin-down electrons. Therefore, they occupy states with different energies: $E_{n\ell}\uparrow$ and $E_{n\ell}\downarrow$, respectively. As a result, a semifilled-subshell atom can be viewed as an atom with fully occupied subshells (the "closed" subshells) which, however, are either completely spin-up or spin-down subshells. Thus, e.g., the ground-state configuration of the Mn atom in SPHF is as follows: Mn(...$3\mathrm{p}^3\uparrow 3\mathrm{p}^3\downarrow 3\mathrm{d}^5\uparrow 4\mathrm{s}^1\uparrow 4s^1\downarrow$, $^6$S).

The choice of SPHF as the zeroth-order basis makes it relatively easy to take into account electron correlation leading to the polarization interaction in a semifilled-shell atomic system. In this paper, to account for electron correlation, we use the random phase approximation with exchange (RPAE), which is built on the basis of SPHF. This approximation is called the spin-polarized RPAE (SPRPAE) [14, 15, 24, 26]. To date, both RPAE and SPRPAE have proven to be effective when applied to closed-shell and semifilled-shell atoms, respectively.

The $e^- + A$ and $e^+ + A$ differential and total scattering cross-sections depend on scattering phases, $\delta_{\ell,s_z}(\epsilon)$ ($\ell$ is the orbital quantum number and $\epsilon$ is the energy of the electron). In SPHF and SPRPAE, $\delta_{\ell,s_z}(\epsilon)$ depends on the electron spin projection, $s_z = \pm\frac{1}{2}$, even in the absence of spin-orbit interaction, that is $\delta_{\ell\uparrow} \neq \delta_{\ell\downarrow}$. This is because of the presence or absence of exchange between the spin-up electrons from a semifilled subshell of the atom and the incoming spin-up and spin-down electrons, respectively. Therefore, the entire $e^- + A$ scattering process for spin-up electrons differs from that for spin-down electrons, and the differences are generally strong [17–19, 21]. In contrast, the $e^+ + A$ scattering does not depend on the positron spin in the absence of spin-orbit interaction. This is due to the absence of exchange between the positron and the electron.

### A. $e^- + A$ scattering

We define the scattering phases of spin-up and spin-down electrons upon their collision with a semifilled shell atom as follows [18, 24, 26]:

$$\delta_\ell^{\uparrow(\downarrow)} = \delta_\ell^{(0)\uparrow(\downarrow)} + \Delta\delta_\ell^{\uparrow(\downarrow)}. \quad (1)$$

Here, $\delta_\ell^{(0)\uparrow(\downarrow)}$ are calculated SPHF scattering phases of the spin-up and spin-down electrons, whereas $\Delta\delta_\ell^{\uparrow(\downarrow)}$ is the correlation (polarization) correction to $\delta_\ell^{(0)\uparrow(\downarrow)}$:

$$e^{i\Delta\delta_{\ell\uparrow(\downarrow)}(\epsilon)} \sin\Delta\delta_{\ell\uparrow(\downarrow)}(\epsilon) = \langle\epsilon\ell_{\uparrow(\downarrow)}|\bar{\Sigma}^\ell(\epsilon)|\epsilon\ell_{\uparrow(\downarrow)}\rangle. \quad (2)$$

In the above equation, $\bar{\Sigma}^\ell(\epsilon)|\epsilon\ell_{\uparrow(\downarrow)}$ is the operator of the reducible self-energy components of the Green-function operator for the incoming electron.

#### 1. *SPRPAE theory in the second-order approximation in the Coulomb interaction (a "$\Sigma^{(2)}$-approximation")*

In the $\Sigma^{(2)}$-approximation, the $\bar{\Sigma}^\ell(\epsilon)$ operator is denoted as $\bar{\Sigma}^{(2)\ell}(\epsilon)$ and presented using the Feynman diagrams in Fig. 2.

Correspondingly, the correlation (polarization) correction, $\Delta\delta_\ell^{\uparrow(\downarrow)}$, is determined as

$$\Delta\delta_{\ell\uparrow(\downarrow)}(\epsilon) = \langle\epsilon\ell_{\uparrow(\downarrow)}|\bar{\Sigma}^{(2)\ell}_{\uparrow(\downarrow)}|\epsilon\ell_{\uparrow(\downarrow)}\rangle. \quad (3)$$

#### 2. *A fuller $\Sigma$-approximation*

If the contribution of the four terms in Fig. 2 is not small, then a fuller account of correlation is in order. Correspondingly, in the fuller SPRPAE approximation, called the $\Sigma$-approximation in the present paper, the $\langle\epsilon\ell_{\uparrow(\downarrow)}|\bar{\Sigma}^\ell_{\uparrow(\downarrow)}|\epsilon\ell_{\uparrow(\downarrow)}\rangle$ matrix element is the solution of the Dyson integral equation with the kernel $\langle\epsilon\ell_{\uparrow(\downarrow)}|\Sigma^\ell_{\uparrow(\downarrow)}|\epsilon'\ell_{\uparrow(\downarrow)}\rangle$ [24, 26]:

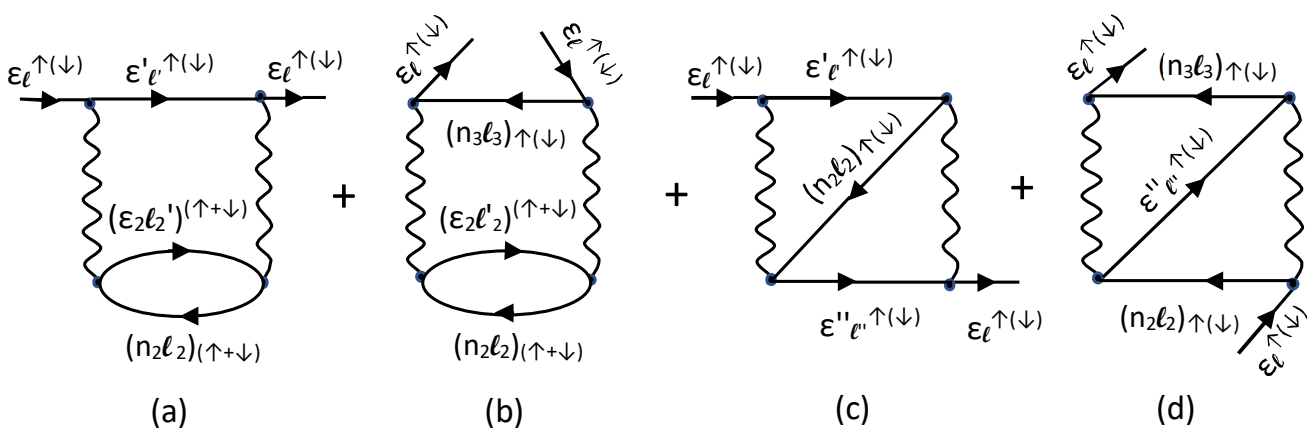

FIG. 2. The reducible self-energy part $\bar{\Sigma}^{(2)\ell}(\epsilon)$ of the Green function of a scattered electron as defined in the SP RPAE theory in the second-order approximation in the Coulomb interaction. Here, a line with a right arrow denotes an electron, a line with a left arrow denotes a vacancy ("hole") in the atom, and a wavy line marks the Coulomb interelectron interaction, $V$. The notations like $\epsilon_\ell$, $\epsilon_{n\ell}$, and $n\ell$, present corresponding states $|\epsilon_l\rangle$, $|\epsilon_{n\ell}\rangle$, and $|n\ell\rangle$ of the electrons and "holes", arrows $\uparrow$ and $\downarrow$ denote the spin-up or spin-down states, respectively. Designation "$(\uparrow + \downarrow)$" means that the contributions of the corresponding spin-up and spin-down states must be summed up in the calculation.

$$\langle \epsilon\ell_{\uparrow(\downarrow)}|\bar{\Sigma}^{\ell}(\epsilon_1)|\epsilon\ell_{\uparrow(\downarrow)}\rangle = \langle \epsilon\ell_{\uparrow(\downarrow)}|\Sigma^{\ell}(\epsilon_1)|\epsilon\ell_{\uparrow(\downarrow)}\rangle + \sum_{\epsilon'} \frac{\langle \epsilon\ell_{\uparrow(\downarrow)}|\Sigma^{\ell}(\epsilon_1)|\epsilon'\ell_{\uparrow(\downarrow)}\rangle\langle \epsilon'\ell_{\uparrow(\downarrow)}|\bar{\Sigma}^{\ell}(\epsilon_1)|\epsilon\ell_{\uparrow(\downarrow)}\rangle}{\epsilon_1 - \epsilon' + i\delta}. \quad (4)$$

Here, the label of summation over the intermediate energy, $\epsilon'$, also assumes integration over the continuous values of $\epsilon'$. Consequently, when the energy of the incident electron, $\epsilon$, exceeds the ionization potential of the perturbed atomic subshell of the atom-scatterer, $\epsilon_1$, the scattering phases become complex: $\delta_\ell^{\uparrow(\downarrow)} = \delta_\ell^{'\uparrow(\downarrow)} + i\delta_\ell^{''\uparrow(\downarrow)}$. Here, $\delta_\ell^{'\uparrow(\downarrow)}$ and $\delta_\ell^{''\uparrow(\downarrow)}$ denote the real and imaginary parts of the scattering phases, respectively.

### B. $e^+ + A$ scattering

Similar to the case of $e^- + A$ scattering, the positron scattering phases, $\delta_\ell^+$, are defined as

$$\delta_\ell^+ = \delta_\ell^{(0)+} + \Delta\delta_\ell^+. \quad (5)$$

Here, $\delta_\ell^{(0)+}$ are SPHF scattering phases of the positron and $\Delta\delta_\ell^+$ is the correlation (polarization) correction to $\delta_\ell^{(0)+}$:

$$e^{i\Delta\delta_\ell^+(\tilde{\epsilon})}\sin\Delta\delta_\ell^+(\tilde{\epsilon}) = \langle \tilde{\epsilon}\tilde{\ell}|\bar{\Sigma}_+^{\ell}(\tilde{\epsilon})|\tilde{\epsilon}\tilde{\ell}\rangle. \quad (6)$$

Here, the tilde sign denotes the positron energy and states and $\bar{\Sigma}_+^\ell(\tilde{\epsilon})$ is the operator of the positron Green function. Similar to $e^- + A$ scattering, we consider $\bar{\Sigma}_+^\ell(\tilde{\epsilon})$ in different approximations. They are listed and explained below.

#### 1. *The $\Sigma_{00}^{(2)}$ and $\Sigma_{00}$ approximations*

These approximations do not take into account the formation of a virtual Ps in $e^+ + A$ scattering. Furthermore, in the $\Sigma_{00}^{(2)}$-approximation, the Green-function operator, $\bar{\Sigma}_{00}^{(2)\ell}(\epsilon)$, is determined only by the diagram shown in Fig. 2(a), where the upper electronic lines must be replaced with the positron lines. In contrast, the $\Sigma_{00}$-approximation is equivalent to the fuller SPRPAE approximation (the $\Sigma$-approximation) for electron scattering, Eq. (4). There, of course, all electron scattering states ($\langle \epsilon\ell|$, etc.) and energies ($\epsilon$, etc.) must be replaced with the positron scattering states and energies, and the electron $\Sigma^{\ell(2)}(\epsilon)$ operator must be replaced by the positron $\bar{\Sigma}_{00}^{(2)\ell}(\epsilon)$ operator.

#### 2. *The $\Sigma_{\rm Ps}^{(2)}$-approximation*

This approximation does take into account the formation of a virtual Ps during the scattering process. The limitation imposed on this approximation is that the Green-function operator, to be denoted as $\bar{\Sigma}_{\rm Ps}^{(2)\ell}(\epsilon)$, is presented only by the diagram shown in Fig. 3, where the shaded oval denotes the interaction of a $e^- + e^+$ virtual pair with other electrons of the atomic core.

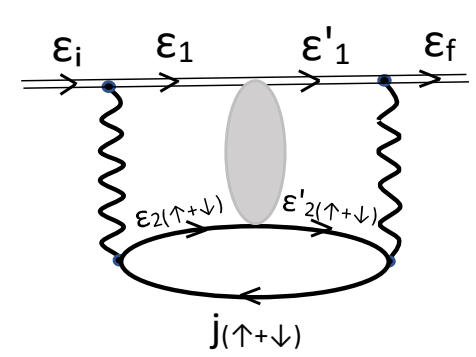

FIG. 3. The irreducible self-energy part $\bar{\Sigma}_{\rm Ps}^{(2)\ell}(\epsilon)$ of the positron Green function in a $\Sigma_{\rm Ps}^{(2)}$-approximation. Here, a double line with a right arrow denotes a positron state, the shaded oval stands for the processes accounting for a virtual $e^- + e^+$ interaction, and other notations carry the same meaning as those in Fig. 2.

In the present paper, as was mentioned in the introduction, we account for the formation of a $e^- + e^+$ virtual pair and its interaction with the rest of the atomic core by means of the subtraction of the magnitude of the $e^- + e^+$ binding energy, $I_{\rm Ps} \approx 6.8$ eV, from the energies of virtual states of the $e^+ + A$ system. Correspondingly, the irreducible matrix element of the $\bar{\Sigma}_{\rm Ps}^{(2)\ell}(\epsilon)$ operator is calculated [26] as

$$\langle\tilde{\epsilon_i}\tilde{\ell_i}||\Sigma^{(2)\ell}_{\rm Ps}(\tilde{\epsilon})||(\tilde{\epsilon_f}\tilde{\ell_f}\rangle = \frac{1}{2\ell+1}\times$$

$$\sum_{j\leq F,L}\sum_{\epsilon_2\ell_2>F,\uparrow(\downarrow)}\int_0^\infty\int_0^\infty \frac{\langle\tilde{\epsilon_i}\tilde{\ell_i},(\epsilon_j\ell_j)_{\uparrow(\downarrow)}||V_L||\tilde{\epsilon_1}\tilde{\ell_1},(\epsilon_2\ell_2)_{\uparrow(\downarrow)}\rangle\langle\tilde{\epsilon_1}\tilde{\ell_1},(\epsilon_2\ell_2)_{\uparrow(\downarrow)}||V_L||\tilde{\epsilon_f}\tilde{\ell_f},(\epsilon_j\ell_j)_{\uparrow(\downarrow)}\rangle}{(2L+1)(\tilde{\epsilon}-\tilde{\epsilon_1}-\epsilon_2+I_{\rm Ps}+\epsilon_j+i\delta)}d\epsilon_2 d\tilde{\epsilon_1}. \qquad (7)$$

Here the "$\leq F$" notation marks the occupied electronic states in the atom, the "$> F$" label denotes the excited discrete and continuous atomic states, and the "tilde" sign denotes positron states.

### 3. The $\Sigma_{\rm Ps}$-approximation

In essence, this approximation is similar to a fuller $\Sigma_{00}$-approximation, but with the important correction: it takes into account the formation of a virtual Ps upon $e^+ + A$ scattering. Correspondingly, in the $\Sigma_{\rm Ps}$-approximation, the matrix element $\langle\tilde{\epsilon}\tilde{\ell}|\bar{\Sigma}^{\ell}_{\rm Ps}(\tilde{\epsilon})|\tilde{\epsilon}\tilde{\ell}\rangle$ in Eq. (6) is the solution of the following Dyson integral equation:

$$\langle\tilde{\epsilon}\tilde{\ell}|\bar{\Sigma}^{\ell}_{\rm Ps}(\epsilon_1)|\tilde{\epsilon}\tilde{\ell}\rangle = \langle\tilde{\epsilon}\tilde{\ell}|\Sigma^{\ell(2)}_{\rm Ps}(\epsilon_1)|\tilde{\epsilon}\tilde{\ell}\rangle + \sum_{\epsilon'}\frac{\langle\tilde{\epsilon}\tilde{\ell}|\Sigma^{\ell(2)}_{\rm Ps}(\epsilon_1)|\tilde{\epsilon}'\tilde{\ell}\rangle\langle\tilde{\epsilon}'\tilde{\ell}|\bar{\Sigma}^{\ell}_{\rm Ps}(\epsilon_1)|\tilde{\epsilon}\tilde{\ell}\rangle}{\epsilon_1-\epsilon'+I_{\rm Ps}+i\delta}. \qquad (8)$$

## III. RESULTS AND DISCUSSION

In the carried out study of $e^+ + {\rm Mn}$ elastic scattering, we accounted for monopole, dipole, quandrupole, and octupole virtual excitations of the three outer subshells of Mn: the $4s\uparrow$, $4s\downarrow$, and $3d\uparrow$ subshells. The contributions of the higher-order multipolar terms and deeper atomic subshells are negligible compared to those indicated above.

### A. Impact of the positron-perturbed $4s\uparrow$ and $4s\downarrow$ subshells on $e^+ + {\rm Mn}$ elastic scattering

Corresponding calculated results for the elastic scattering phase shifts of the $s$, $p$, $d$, and $f$ partial positronic waves in the case of $e^+ + {\rm Mn}$ scattering, obtained in various approximations, are depicted in Fig. 4.

One can see that calculated results, obtained without ("$\Sigma_{00,4s}$-line") and with ("$\Sigma_{\rm Ps,4s}$-line") taking into account the formation of a virtual Ps in the calculations of $\delta_\ell$s, differ from each other dramatically. Thus, the influence of the formation of a virtual Ps on the scattering process is so great that it cannot be excluded from the study.

Next, note how the impact of electron correlation on scattering phases depends on a utilized approximation.

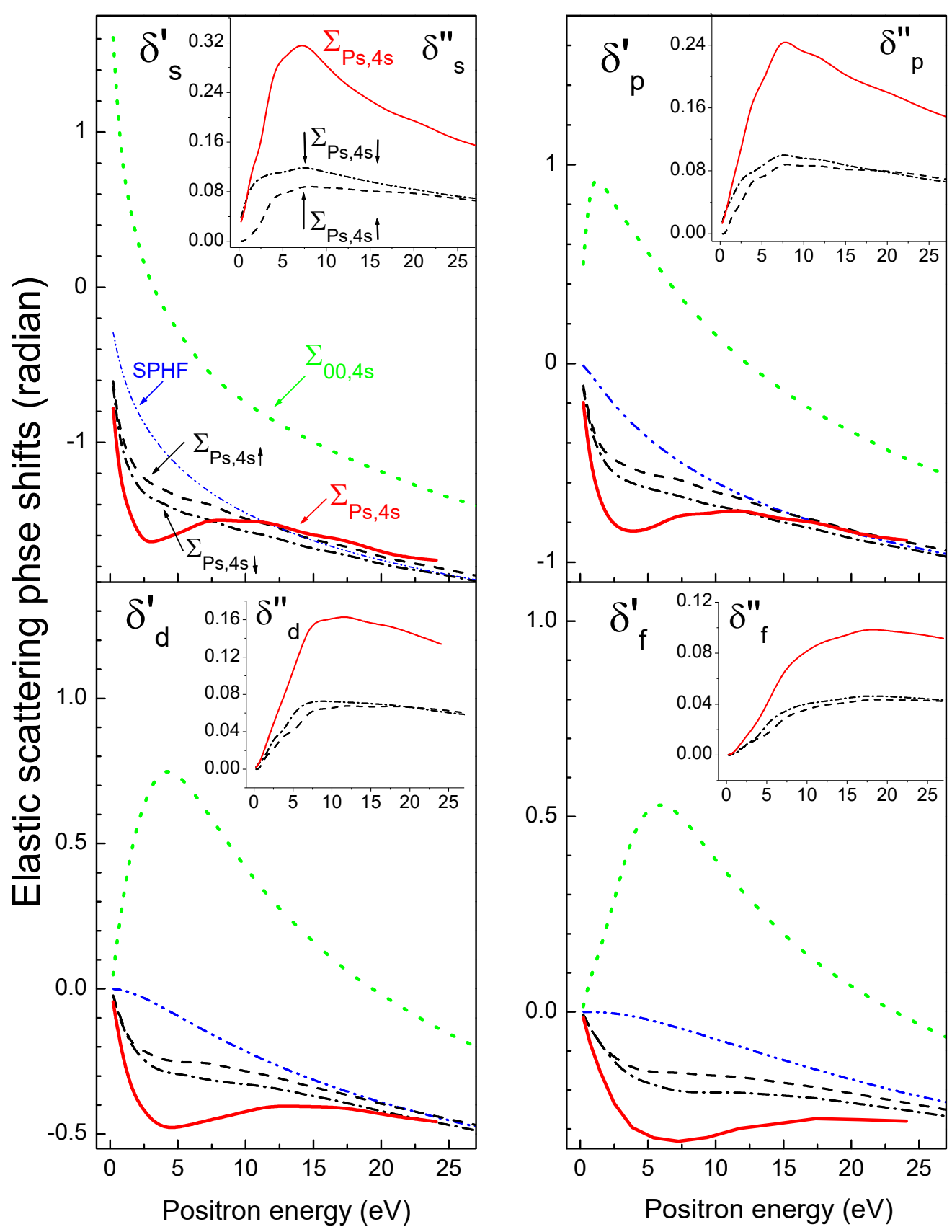

FIG. 4. (Color online) Real ($\delta'_\ell$) and imaginary ($\delta''_\ell$) parts of the elastic scattering phase shifts, $\delta_\ell$ (in the units of radian), in the case of $e^+ + {\rm Mn}$ collision, as marked. The corresponding calculations took into account the individual responses of the Mn $4s\uparrow$-subshell [dash, ($\Sigma_{\rm Ps,4s\uparrow}$)] and the $4s\downarrow$-subshell [dash-dot, ($\Sigma_{\rm Ps,4s\downarrow}$)], the collective response of the $4s\uparrow + 4s\downarrow$ subshells [solid, ($\Sigma_{\rm Ps,4s}$)] all being calculated in the $\Sigma_{\rm Ps}$-approximation, as well as the collective response of the $4s\uparrow + 4s\downarrow$ subshells calculated both in the $\Sigma_{00}$-approximation [dots, ($\Sigma_{00,4s}$)] and SPHF (dash-dot-dot).

Specifically, the $\Sigma_{00,4s}$-approximation predicts a very strong effect of electron correlation in the entire energy region (cf. the "SPHF" and "$\Sigma_{00,4s}$" lines in Fig. 4). However, the more complete $\Sigma_{\rm Ps,4s}$-approximation, which takes into account the formation of a virtual Ps in the scattering process, shows that the correlation effect is strong only up to about 10 eV of the positron energy (cf. the "SPHF" and "$\Sigma_{\rm Ps,4s}$" lines in the figure).

Now, let us discuss, on a relative scale, the individual and aggregated impacts, exerted by the perturbed $4s\uparrow$ and $4s\downarrow$ subshells, on the scattering phases. To do this, intercompare graphs "$\Sigma_{\mathrm{Ps},4s\uparrow}$" (dash), "$\Sigma_{\mathrm{Ps},4s\downarrow}$" (dash-dot), and "$\Sigma_{\mathrm{Ps},4s}$" (solid) in Fig. 4. It is seen that the individual effects of these subshells on scattering phases (cf. the "$\Sigma_{\mathrm{Ps},4s\uparrow}$" and "$\Sigma_{\mathrm{Ps},4s\downarrow}$" lines) differ from each other. It was expected, of course. However, the relative difference between them is not strong, except, perhaps, for the difference between the imaginary parts of the scattering phases of the $s$-positron-wave at low incident energies. More interesting and important is that the energy dependence of the real parts of the phase shifts, calculated with taking into account only the individual impact of the $4s\uparrow$ or $4s\downarrow$ subshell on scattering phases, is close to a monotonic energy dependence, with insignificant exceptions, which are most noticeable for the $f$-wave. In contrast, the aggregated influence of these subshells on scattering phases results in a pronouncedly non-monotonic energy dependence of the phases in the energy region up to about 10 eV. This reveals strong interference between the individual impacts of the $4s\uparrow$ or $4s\downarrow$ subshell on the scattering phases.

Finally, we note that the influence of both electron correlation and the formation of a virtual Ps on the scattering phases turns out to be especially significant, both quantitatively and qualitatively, at low scattering energies.

### B. Impact of the positron-perturbed $3d^5\uparrow$ subshell on $e^+ + \mathrm{Mn}$ elastic scattering

Corresponding calculated results for the scattering phases for the $s$, $p$, $d$, and $f$ partial positron waves, obtained in the SPHF, $\Sigma_{00,3d}$, and $\Sigma_{\mathrm{Ps,3d}}$ approximations, are depicted in Fig. 5.

One can see that the relative difference between the real parts of the scattering phases, calculated in the $\Sigma_{00,3d}$ and $\Sigma_{\mathrm{Ps,3d}}$ approximations, is not strong. In other words, the effect of the formation of a virtual Ps on the scattering phases in this case is relatively weak, in comparison with this effect in the case of the aggregated impact of the $4s\uparrow$ and $4s\downarrow$ subshells on the scattering phases (Fig. 4). In our opinion, this is due to the "inner-orbital"-side of the $3d^5\uparrow$ subshell whose mean radius is about only 1 a.u.

Neither is strong the effect of electron correlation on the scattering phases (cf. graphs "$\Sigma_{00,3d}$", "$\Sigma_{\mathrm{Ps,3d}}$", and "SPHF" in Fig. 5) in comparison to that owing to the combined influence of the $4s\uparrow$ and $4s\downarrow$ subshells on the scattering phases. Nevertheless, the electron correlation influence of the $3d^5\uparrow$-subshell on the scattering phases is stronger than the influence of the formation of a virtual Ps in the excited spectrum of this subshell on the scattering process. We attribute the stronger role of electron correlation in this case to the "valence-orbital"-side of the $3d^5\uparrow$-subshell nature which is reflected in its weak binding energy and large dynamical polarizability.

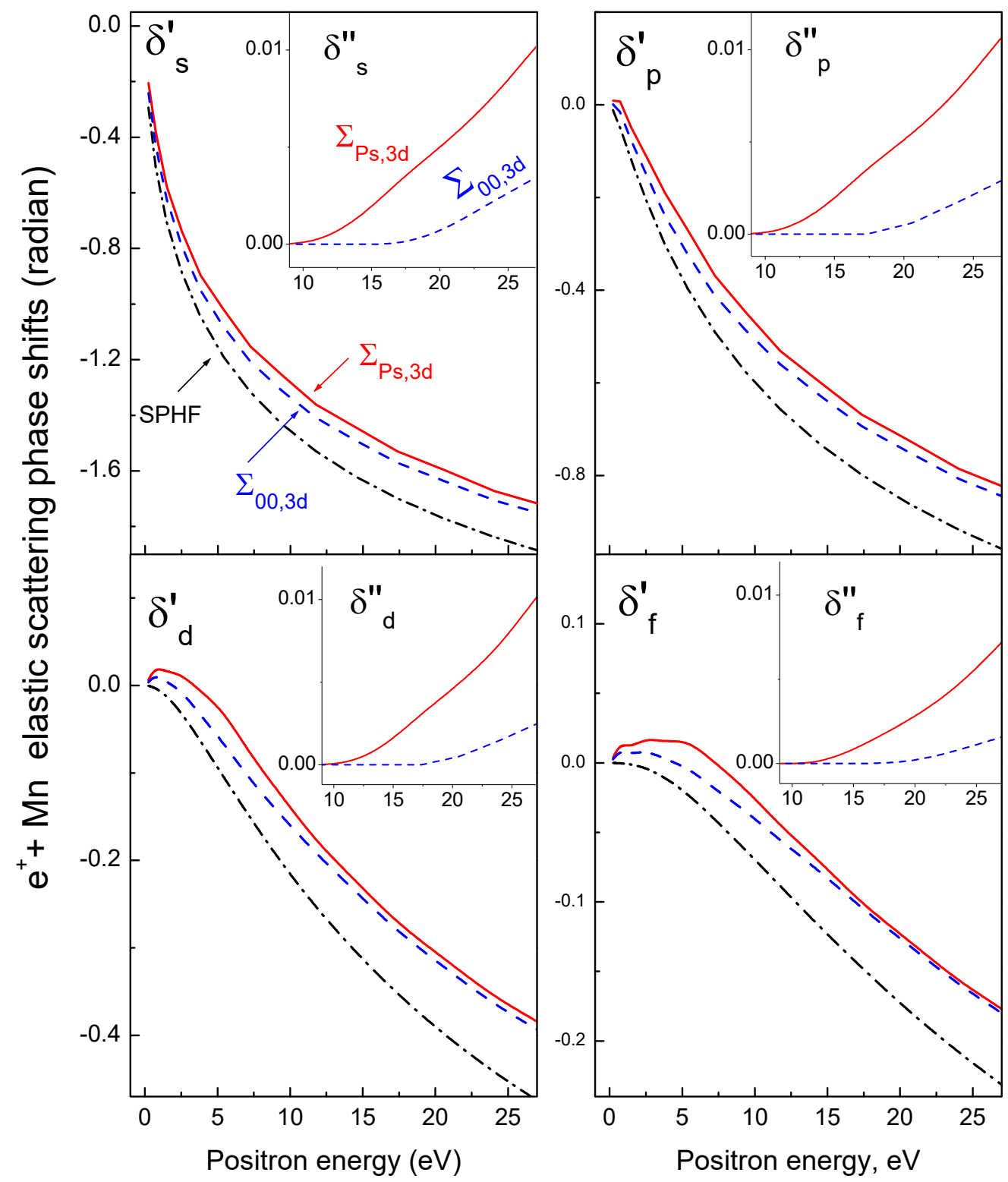

FIG. 5. (Color online) Real ($\delta'_\ell$) and imaginary ($\delta''_\ell$) parts of the $e^+ + \mathrm{Mn}$ elastic scattering phase shifts, $\delta_\ell$ calculated in SPHF (dash-dot) as well as in the $\Sigma_{00,3d}$ (dash) and $\Sigma_{\mathrm{Ps,3d}}$ (solid) approximations accounting for only the individual correlated impact of the $3d^5\uparrow$ subshell on the scattering process.

It is interesting to note that all of the above is in a sharp contrast to a much strong influence of both the electron correlation and formation of a virtual Ps on the scattering phases in the case of the combined impact of the $4s\uparrow$ and $4s\downarrow$ subshells on the scattering process.

### C. Combined effect of the $4s\uparrow$, $4s\downarrow$, and $3d^5\uparrow$ subshells on $e^+ + \mathrm{Mn}$ elastic scattering

The combined correlated influence of the positron-perturbed $4s\uparrow$, $4s\downarrow$, and $3d^5\uparrow$ subshells on the scattering phases was calculated in two approximations, namely, without taking into account the formation of a virtual Ps in the scattering process (the $\Sigma_{00,4s+3d}$ approximation) and with this formation taken into account (the $\Sigma_{Ps,4s+3d}$ approximation). The calculated data, together with the scattering phases calculated in the $\Sigma_{00,4s}$ and $\Sigma_{Ps,4s}$ approximations (see Fig. 4), are shown in Fig. 6.

Intercomparison of the presented results reveals a profoundly interesting feature. Indeed, let us recall that the individual influence of the positron-perturbed $3d^5\uparrow$ semi-filled subshell of Mn on scattering phases is relatively small both on an absolute scale (Fig. 5) and in comparison with the effect associated only with the $4s\uparrow$ and $4s\downarrow$ subshells (see Fig. 4). However, when perturbation of all three subshells - $4s\uparrow$, $4s\downarrow$, and $3d^5\uparrow$ - by the incoming positron is taken into account, the results obtained (graphs "$\Sigma_{Ps,4s+3d}$" in Fig. 6) differ drastically from the results obtained in the $\Sigma_{Ps,4s}$ approximation alone (graphs "$\Sigma_{Ps,4s}$" in Fig. 6). The differences, both quantitative and qualitative, are particularly spectacular between the real parts of the scattering phases calculated in the $\Sigma_{Ps,4s+3d}$ and $\Sigma_{Ps,4s}$ approximations. We attribute the discovered peculiarity in the scattering phases calculated in the $\Sigma_{Ps,4s+3d}$ approximation to the "valence-orbital"-side of the $3d^5\uparrow$-subshell nature. Namely, because its binding energy ($\approx 14$ eV) is relatively close to the binding energies of the $4s\uparrow$ and $4s\downarrow$ subshells ($\approx 8$ eV), it is as easily perturbed by the incoming positron as the single-electron $4s\uparrow$ and $4s\downarrow$ subshells. However, its oscillator strength significantly exceeds the oscillator strengths of the $4s\uparrow$ and $4s\downarrow$ subshells. Correspondingly, the excitation channels of the $4s\uparrow$ and $4s\downarrow$ subshells are strongly influenced by the excitation channels of the $3d^5\uparrow$ subshell due to interchannel coupling. Therefore, the combined impact of the $4s\uparrow$, $4s\downarrow$, and $3d^5\uparrow$ subshells on the scattering of the positron by the atom turns out to be stronger than the combined influence of only the $4s\uparrow$ and $4s\downarrow$ subshells. This happens even though the individual influence of the $3d^5\uparrow$ subshell alone on the scattering process is by itself relatively weak.

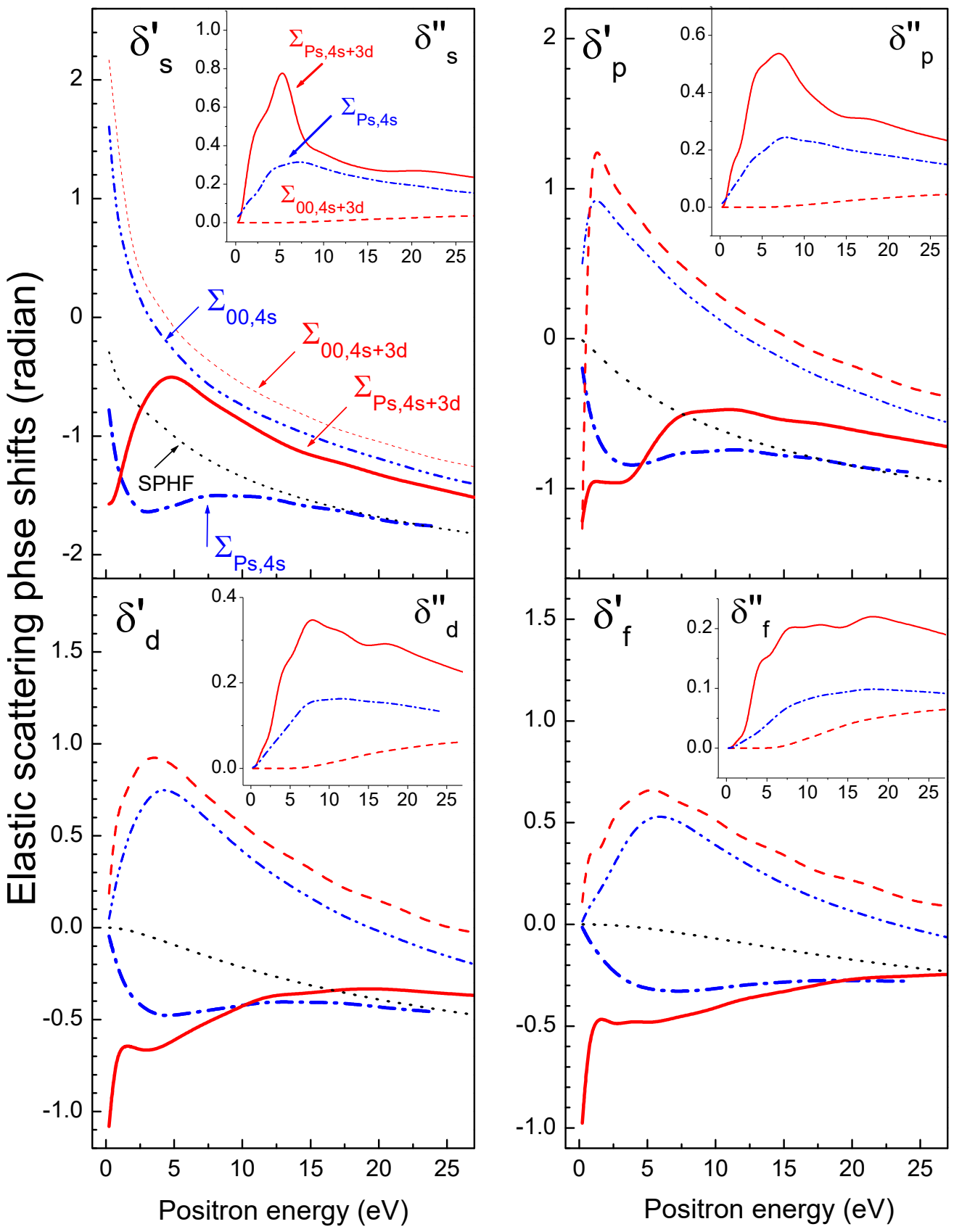

FIG. 6. (Color online) Real ($\delta'_\ell$) and imaginary ($\delta''_\ell$) parts of the elastic scattering phases ($\delta_\ell$) calculated in various approximations for the case of $e^+ +$Mn scattering. Dots, SPHF. Dash, $\Sigma_{00,4s+3d}$ (see text). Solid, $\Sigma_{Ps,4s+3d}$ (see text). Dash-dot-dot, $\Sigma_{00,4s}$ (see Fig. 4). Dash-dot, $\Sigma_{4s}$ (see Fig. 4).

In the next, we discuss the $e^+ +$Mn total elastic scattering cross section, calculated by taking into account the combined influence of, (a), only the positron-perturbed $4s\uparrow$ and $4s\downarrow$ subshells in the $\Sigma^{(2)}_{\mathrm{Ps,4s}}$ and $\Sigma_{\mathrm{Ps,4s}}$ approximations and, (b), the positron-perturbed $4s\uparrow$, $4s\downarrow$, and $3d^5\uparrow$ subshells in the $\Sigma^{(2)}_{\mathrm{Ps,4s+3d}}$ and $\Sigma_{\mathrm{Ps,4s+3d}}$ approximations. The corresponding calculated data are plotted in Fig. 7.

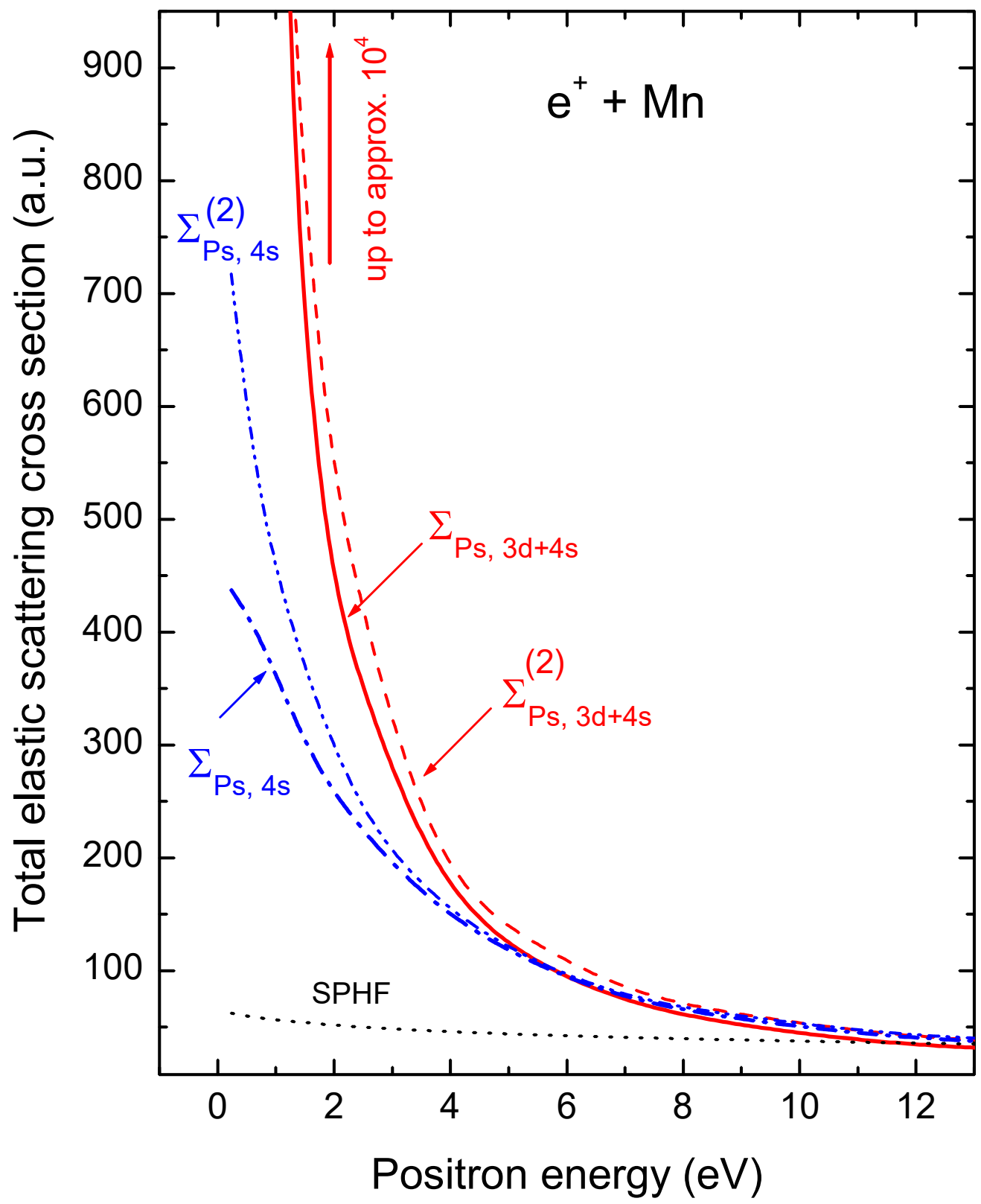

FIG. 7. (Color online) The total elastic scattering cross section in the case of the $e^+ +$Mn collision calculated in SPHF and other approximations by taking into account the combined influence of only two $4s\uparrow$ and $4s\downarrow$ subshells on the scattering process (the $\Sigma^{(2)}_{\mathrm{Ps,4s}}$ and $\Sigma_{\mathrm{Ps,4s}}$ approximations) as well as the combined impact of the three $4s\uparrow$, $4s\downarrow$, and $3d^5\uparrow$ subshells (the $\Sigma^{(2)}_{\mathrm{Ps,4s+3d}}$ and $\Sigma_{\mathrm{Ps,4s+3d}}$ approximations), as marked in the figure.

Perhaps, the most unusual result is that the inclusion of the influence of the "inner-valence" $3d^5\uparrow$ subshell in addition to that of the two outer $4s\uparrow$ and $4s\downarrow$ subshells radically changes the numerical value of the scattering cross section with decreasing energy of the incident positron in comparison with the results obtained in the $\Sigma^{(2)}_{\mathrm{Ps,4s}}$ and $\Sigma_{\mathrm{Ps,4s}}$ approximations. It is also interesting to note that calculated data obtained in the two latter approximations differ significantly from each other at low positron energies, whereas the calculated results obtained in the $\Sigma^{(2)}_{\mathrm{Ps,4s+3d}}$ and $\Sigma_{\mathrm{Ps,4s+3d}}$ approximations stay close to each other in the entire energy domain.

### D. $e^+ + \mathrm{Mn}$ scattering versus $e^- + \mathrm{Mn}$

Finally, in Fig. 8 we demonstrate the differences between the calculated $e^+ + \mathrm{Mn}$ elastic scattering cross section, $\sigma^+$, and our earlier calculated [18] spin-averaged $e^- + \mathrm{Mn}$ elastic scattering cross section, $\sigma^-$, obtained in the framework of the identical formalism.

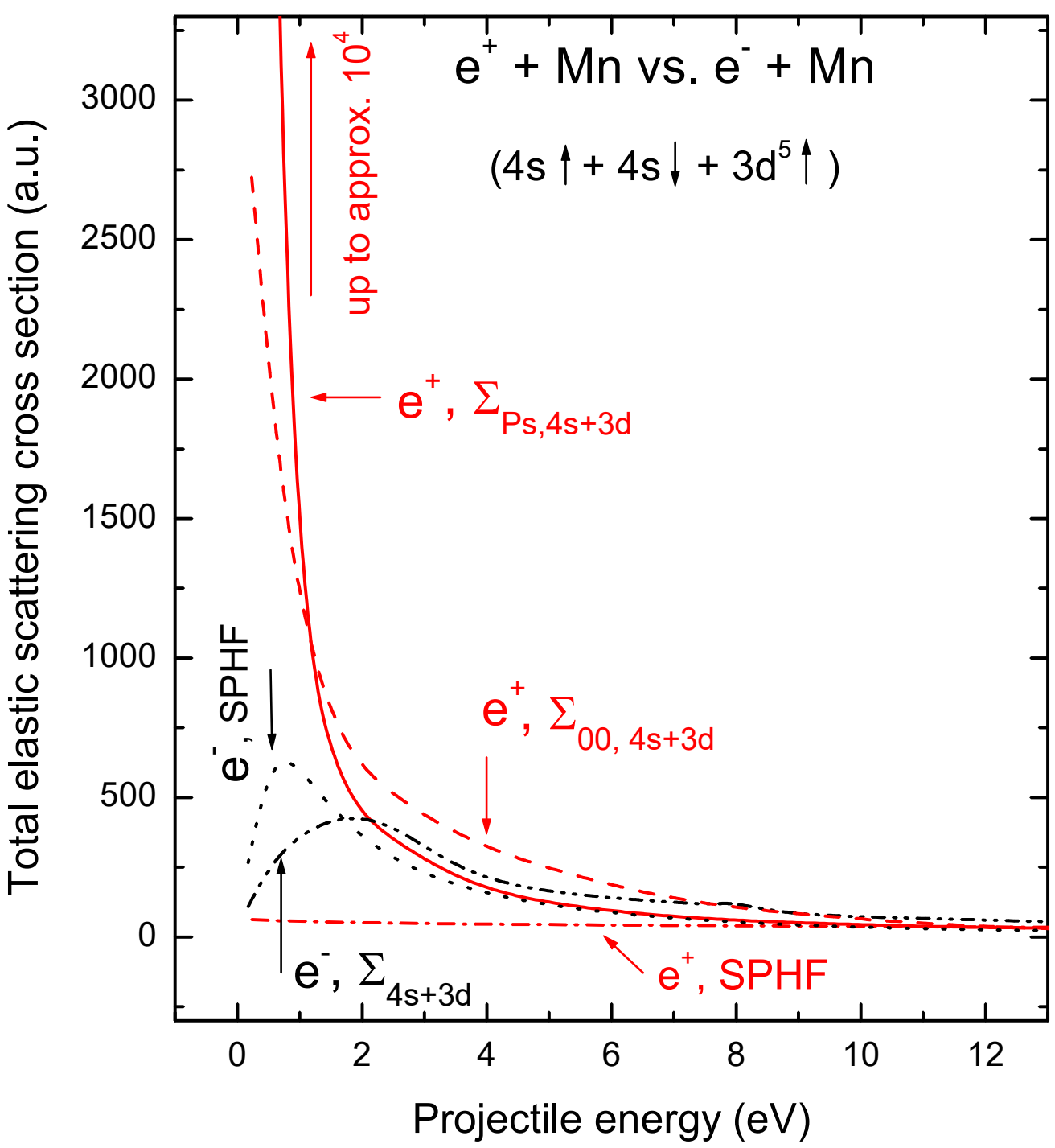

FIG. 8. (Color online) Total $e^+ + \mathrm{Mn}$ elastic scattering cross section, $\sigma^+$, versus the spin-averaged $e^- + \mathrm{Mn}$ elastic scattering cross section, $\sigma^-$ [18]. In both cases, the calculations included the combined influence of the $4s\uparrow$, $4s\downarrow$, and $3d^5\uparrow$ subshells on the scattering process. Dots, $\sigma^-$ calculated in SPHF. Dash-dot-dot, $\sigma^-$ calculated in the $\Sigma_{\mathrm{4s+3d}}$ approximation. Dash-dot, $\sigma^+$ calculated in SPHF. Dash, $\sigma^+$ calculated in the $\Sigma_{\mathrm{00,4s+3d}}$ approximation. Solid, $\sigma^+$ calculated in the $\Sigma_{\mathrm{Ps,4s+3d}}$ approximation.

Calculated data show significant quantitative and qualitative differences between $\sigma^+$ and $\sigma^-$ at low projectile energies regardless of the approximation used. For instance, the energy dependence of $\sigma^+$ and $\sigma^-$, calculated in the corresponding $\Sigma$-approximations, is seen to be taking qualitatively opposite routes when the projectile energy falls below about 2 eV. Moreover, $\sigma^+$ begins to absolutely dominate $\sigma^-$ with decreasing energy. We believe that we have discovered a unique situation where the positron-atom scattering dominates the electron-atom scattering to a degree that has not been observed in any earlier studies of the $e^+ + A$ and $e^- + A$ scattering processes.

Taking the opportunity presented by Fig. 8, we emphasize once again the importance of accounting for the formation of a virtual Ps during the $e^+ + \mathrm{Mn}$ scattering process (the $\Sigma_{\mathrm{Ps,4s+3d}}$ approximation) that hugely alters the cross section calculated without this formation taken into account (the $\Sigma_{\mathrm{00,4s+3d}}$ approximation).

## IV. CONCLUSION

In conclusion, we have provided the insight into elastic scattering of positrons by an atom with a multielectron semifilled subshell - the $\mathrm{Mn}(...3d^5 4s^2, {}^6S)$ atom.

One particularly surprising finding is that, although the individual influence of the "inner-outer" $3d^5\uparrow$ semifilled subshell of Mn on $e^+ + \mathrm{Mn}$ scattering is not strong, but, when its virtual excitation is combined with virtual excitations of the $4s\uparrow$ and $4s\downarrow$ subshells, the net effect turns out to be significantly stronger than the effect caused by the $4s\uparrow$ and $4s\downarrow$ subshells alone on the scattering process.

Another spectacular finding is that the effect of electron correlation on $e^+ + \mathrm{Mn}$ scattering, with decreasing energy of an incident particle, takes the opposite route compared to the case of $e^- + \mathrm{Mn}$ scattering. On the whole, the $e^+ + \mathrm{Mn}$ elastic scattering cross section proves to be dominant over the $e^- + \mathrm{Mn}$ elastic scattering cross section at low energies up to a degree not previously observed in the studies of the scattering of positrons and electrons by atoms of alkali metals and noble gases. The rise of the $e^+ + \mathrm{Mn}$ scattering cross section suggests the existence of a weakly bound positron state near a zero energy. On the other hand, it is known that Mn cannot form a negative ion, so that no near-zero bound states of an electron in the field of Mn can exist. As a result, one could expect the formation of near-threshold resonances in elastic scattering of positrons off Mn in contrast to the case of the electron-Mn scattering, and our study has clearly identified the actual existence of such resonance in $e^+ + \mathrm{Mn}$ scattering and its natural absence in $e^- + \mathrm{Mn}$ scattering. It is because of this qualitative difference between the $e^+ + \mathrm{Mn}$ and $e^- + \mathrm{Mn}$ scattering that the elastic scattering cross section takes an opposite route for $e^+ + \mathrm{Mn}$ scattering compared to $e^- + \mathrm{Mn}$ at a near-zero energy.

And, of course, it has been undoubtedly demonstrated in the present paper that taking into account both the electron correlation and the formation of a virtual Ps in the calculation of $e^+ + \mathrm{Mn}$ elastic scattering is crucial for understanding the scattering process.

We hope that our study has laid the ground for both future theoretical and experimental studies of the scattering of positrons by the semifilled-shell atoms.

[1] M. Ya. Amusia, N. A. Cherepkov, L. V. Chernysheva, and S. G. Shapiro, Elastic scattering of slow positrons by helium, J. Phys. B, **9**, L531 (1976). https://iopscience.iop.org/article/10.1088/0022-3700/9/17/005

[2] V. A. Dzuba, V. V. Flambaum, G. F. Gribakin, and W. A. King, Many-body calculations of positron scattering and annihilation from noble-gas atoms, J. Phys. B **29**, 3151 (1996). DOI: 10.1088/0953-4075/29/14/024

[3] G. F. Gribakin and W. A. King, Positron scattering from Mg atom, Can. J. Phys. **74**, 449 (1996). https://doi.org/10.1139/p96-064.

[4] M. Ya. Amusia, N. A. Cherepkov, and L. V. Chernysheva, Elastic scattering of slow positrons on atoms, JETP **97**, 34 (2003). DOI:10.1134/1.1600794

[5] C. M. Surko, G. F. Gribakin, and S. J. Buckman, Low-energy positron interactions with atoms and molecules, J. Phys. B **38** R57 (2005). https://iopscience.iop.org/article/10.1088/0953-4075/38/6/R01

[6] V. A. Dzuba, V. V. Flambaum, and G. F. Gribakin, Detecting Positron-Atom Bound States through Resonant Annihilation, Phys. Rev. Lett. **105**, 203401 (2010). https://doi.org/10.1103/PhysRevLett.105.203401

[7] A. C. L. Jones, C. Makochekanwa, P. Caradonna, *et.al*, Positron scattering from neon and argon, Phys. Rev. A **83**, 032701 (2011). DOI: https://doi.org/10.1103/PhysRevA.83.032701.

[8] D. G. Green, J. A. Ludlow, and G. F. Gribakin, Positron scattering and annihilation on noble-gas atoms, Phys. Rev. A **90**, 032712 (2014). DOI: https://doi.org/10.1103/PhysRevA.90.032712.

[9] L. Chiari and A. Zecca, Recent positron-atom cross section measurements and calculations, Eur. Phys. J. D **68** 297 (2014). https://doi.org/10.1140/epjd/e2014-50436-4

[10] A. S. Kadyrov and I. Bray, Recent progress in the description of positron scattering from atoms using the convergent close-coupling theory, J. Phys. B **49**, 222002 (2016). https://iopscience.iop.org/article/10.1088/0953-4075/49/22/222002.

[11] S. N. Nahar and B. Antony, Positron Scattering from Atoms and Molecules, Atoms, **8**, No. 29 (30pp), 2020. doi:10.3390/atoms8020029.

[12] W. Williams, J. C. Cheesborough, and S. Trajmar, Elastic and inelastic scattering of electrons by atomic manganese, J. Phys. B **11**, 2031 (1978). https://iopscience.iop.org/article/10.1088/0022-3700/11/11/019.

[13] R. Meintrup, G. F. Hanne, and K. Bartschat, Spin exchange in elastic collisions of polarized electrons with manganese atoms, J. Phys. B **33**, L289 (2000). https://iopscience.iop.org/article/10.1088/0953-4075/33/8/102.

[14] M. Ya. Amusia, V. K. Dolmatov, and V. K. Ivanov, Photoionization of atoms with half-filled shells, Zh. Eksp. Teor. Fiz. **85**, 115 (1983) [Sov. Phys. JETP **58**, 67 (1983)].

[15] M. Ya. Amusia and V. K. Dolmatov, Photoionization of inner $ns$ electrons in semifilled shell atoms ($3s$ electrons in a Mn atom), J. Phys. B **26** 1425 (1993). https://iopscience.iop.org/article/10.1088/0953-4075/26/8/010.

[16] V. K. Dolmatov, A. S. Kheifets, P. C. Deshmukh, and S. T. Manson, Attosecond time delay in the photoionization of Mn in the region of the $3p \to 3d$ giant resonance, Phys. Rev. A **91**, 053415 (2015). DOI: 10.1103/PhysRevA.91.053415.

[17] M. Ya. Amusia and V. K. Dolmatov, Elastic scattering of electrons by atoms with half-filled subshells, Zh. Eksp. Teor. Fiz. **97**, 1129 (1990) [Sov. Phys. JETP **70**, 632 (1990)].

[18] V. K. Dolmatov, M. Ya. Amusia, and L. V. Chernysheva, Electron elastic scattering off a semifilled-shell atom: The Mn atom, Phys. Rev. A **88**, 042706 (2013). DOI: 10.1103/PhysRevA.88.042706

[19] V. K. Dolmatov, Angle-differential elastic electron scattering off Mn, Phys. Rev. A **96**, 052704 (2017). DOI: 10.1103/PhysRevA.96.052704.

[20] V. K. Dolmatov, Characteristic features of the $3p$ absorption spectra of free iron-group elements due to the duplicity of the "inner-valence" $3d$ electrons. Application to $Mn^{2+}$, J. Phys. B **29**, L687 (1996). https://iopscience.iop.org/article/10.1088/0953-4075/29/19/001.

[21] V. K. Dolmatov, M. Ya. Amusia, and L. V. Chernysheva, Electron elastic scattering off a spin-polarized Cr atom, Phys. Rev. A **90**, 032717 (2014). DOI: 10.1103/PhysRevA.90.032717.

[22] M. S. Dababneh, W. E. Kauppila, J. P. Downing, F. Laperriere, V. Pol, J. H. Smart, and T. S. Stein, Measurements of total scattering cross sections for low-energy positrons and electrons colliding with krypton and xenon, Phys. Rev. A **22** 1872 (1978). https://doi.org/10.1103/PhysRevA.22.1872

[23] G. Sinapius, W. Raith, and W. G. Wilson, Scattering of low-energy positrons from noble-gas atoms, J. Phys. B **13** 4079 (1980). https://iopscience.iop.org/article/10.1088/0022-3700/13/20/020

[24] M. Ya. Amusia and L. V. Chernysheva, *Computation of Atomic Processes: A Handbook for the ATOM Programs* (IOP, Bristol, 1997).

[25] J. C. Slater, *The Self-Consistent Field for Molecules and Solids* (McGraw-Hill, New York, 1974).

[26] M. Ya. Amusia, L. V. Chernysheva, and V. G. Yarzhemsky, Handbook of theoretical Atomic Physics, Data for photon absorption, electron scattering, vacancies decay (Springer, Berlin, 2012).